\documentclass[twocolumn,showpacs,preprintnumbers,amsmath,amssymb,prl]{revtex4}

\usepackage{graphicx}
\usepackage{dcolumn}
\usepackage{bm}

\begin{document}

\title{Fermi surface topology and the upper critical field 
in two-band superconductors - application to MgB$_2$}

\author{T.~Dahm and N.~Schopohl}

\affiliation{Institut f\"ur Theoretische Physik, 
         Universit\"at T\"ubingen, 
         Auf der Morgenstelle 14, D-72076 T\"ubingen, 
         Germany}

\date{\today}

\begin{abstract}
Recent measurements of the anisotropy of the upper critical field $B_{c2}$
on MgB$_2$ single crystals have shown a puzzling strong temperature
dependence. Here, we present a calculation of the upper critical field
based on a detailed modeling of bandstructure calculations that takes
into account both the unusual Fermi surface topology and the two gap
nature of the superconducting order parameter. Our results show that the
strong temperature dependence of the $B_{c2}$ anisotropy can be understood
as an interplay of the dominating gap on the $\sigma$-band, which possesses
a small
$c$-axis component of the Fermi velocity, with the induced superconductivity 
on the $\pi$-band possessing a large $c$-axis component of the Fermi velocity. 
We provide analytic formulas for the
anisotropy ratio at $T=0$ and $T=T_c$ and quantitatively predict the 
distortion of the vortex lattice based on our calculations.
\end{abstract}


\maketitle


Our understanding of the physical properties of the recently discovered
superconductivity in MgB$_2$ has made rapid progress since its discovery
\cite{Akimitsu}. Its high critical temperature $T_c=39$~K can be
understood as arising from strong conventional electron-phonon coupling
to a high frequency phonon mode \cite{Dolgov,Liu,Heid}. 
Its pairing symmetry seems to be of
conventional $s$-wave type \cite{Manzano,Yang}. However, in contrast to 
conventional superconductors, a number of recent experiments
indicate that there exist two gaps of different size in this compound
\cite{Szabo,Giubileo,Bouquet,Sologubenko,Bouquet2,Iavarone}.
This possibility is supported by band structure calculations, which have 
shown that the Fermi surface of this compound consists of four bands: two
$\sigma$-type two-dimensional cylindrical hole sheets and two $\pi$-type
three dimensional tubular networks \cite{Shulga,Kortus} in good overall
agreement with recent de Haas-van Alphen experiments \cite{Carrington}. 
Microscopic calculations of the superconducting gap
based on band structure calculations have shown recently that indeed
one should expect a big superconducting gap living on the $\sigma$ bands
and a smaller one, induced by interband pairing interaction, living on the
$\pi$ bands \cite{Liu,Louie}. Impurity scattering, which in conventional 
superconductors
tends to average out strongly differing gap values, in this case
becomes ineffective, because the $\sigma$ and $\pi$ bands possess
different symmetries, making interband impurity scattering much
weaker than intraband impurity scattering \cite{Mazin}.

Recent measurements of the upper critical field $B_{c2}$, particularly its
anisotropy, on single crystal MgB$_2$ have shown a puzzling strong
temperature dependence of the anisotropy ratio $B_{c2}^{ab}/B_{c2}^c$ between
the ab-plane and the c-axis upper critical field \cite{Angst,Eltsev,Lyard}. 
In conventional
systems this ratio rarely changes by more than 10 to 20 percent as a function of
temperature. In MgB$_2$ changes by more than a factor of 2 have been observed.
It has been shown that such a strong temperature dependence of the
anisotropy ratio can be obtained within a strongly anisotropic single gap
model, possessing a small gap in c-axis direction and a big gap in
ab-plane direction \cite{Posa1}. However, the gap anisotropy would have to be a factor
of 10, which is too big as compared with experimental values. In addition,
this scenario would be inconsistent with penetration depth studies
which clearly indicate the presence of a small gap within the ab-plane
\cite{Jin,Posa2}.

Here, we want to present first calculations of the upper critical field
$B_{c2}$, which take into account the multi-band Fermi surface 
structure seriously. We show that the strong temperature dependence
of the anisotropy of $B_{c2}$ can be traced back to the influence of
the two topologically very different Fermi surface types. While the cylindrical
Fermi surface sheets are dominating the behavior of $B_{c2}$ at low
temperatures, leading to a large anisotropy, at temperatures approaching
$T_c$ the $\pi$ bands due to their much larger c-axis Fermi velocity start
to play a more important role, strongly reducing the $B_{c2}$ anisotropy. Thus,
the strong temperature dependence of the $B_{c2}$ anisotropy appears
as a cross-over from a low temperature $\sigma$ band dominated regime
to a higher temperature mixed $\sigma$ and $\pi$ band regime.

In order to include the multi-band
Fermi surface structure in the calculation of the upper critical field,
we start our investigation from the fully momentum dependent multi-band
formulation of the quasiclassical (Eilenberger) theory of the 
upper critical field
\cite{Rieck}. For that purpose we have to solve the linearized multi-band 
gap equation in the presence of an external magnetic field, which reads
\begin{equation}
\Delta_\alpha (\vec{r}) = - \pi T
\sum_{\alpha'} \sum_{|\omega_n^\prime| < \omega_c} \lambda^{\alpha \alpha'} 
\left\langle
f_{\alpha'} (\vec{r}, \hat{k}^\prime; \omega_n^\prime)
\right\rangle_{\alpha'}
\label{eq1}
\end{equation}
Here, $f_{\alpha}$ is the anomalous Eilenberger propagator on the Fermi
surface sheet denoted by $\alpha$. $\Delta_\alpha$ is the gap function
on that Fermi surface sheet and $\lambda^{\alpha \alpha'}$ is the
pairing interaction, which becomes a matrix in the band indices. The
brackets $\langle \cdots \rangle_{\alpha'}$ denote a Fermi surface average
over momentum $\hat{k}^\prime$ of the Fermi surface sheet $\alpha'$. 
In Eq. (\ref{eq1}) we have already
assumed that the gaps are isotropic $s$-wave on the Fermi surfaces, as
indicated by experiment, but may have different values on different sheets.
At $B_{c2}$ the anomalous Eilenberger propagator has to be determined from the
linearized Eilenberger equation (for $\omega_n>0$)
\begin{equation}
\left\{ \omega_n +
\vec{v}_{F,\alpha} \left[
\frac{\hbar}{2} \vec{\nabla} - i \frac{e}{c} \vec{A} \left( \vec{r} \right)
\right] \right\}  f_{\alpha} (\vec{r}, \hat{k}; \omega_n) =
- \Delta_\alpha (\vec{r})
\label{eq2} 
\end{equation}
Here, $\vec{v}_{F,\alpha}$ is the (momentum dependent)
Fermi velocity on Fermi surface sheet $\alpha$ and $\vec{A}$ the vector
potential due to the magnetic field $\vec{B}=\vec{\nabla} \times \vec{A}$.
Eqs. (\ref{eq1}) and (\ref{eq2}) constitute an eigenvalue problem for 
$\Delta_\alpha (\vec{r})$ . For a given temperature $T$ the solution 
$\Delta_\alpha (\vec{r})$ which solves Eqs. (\ref{eq1})  and (\ref{eq2})
for the highest value of $B$ determines $B_{c2}$. 

\begin{figure}
  \begin{center}
    \includegraphics[width=0.75\columnwidth,angle=0]{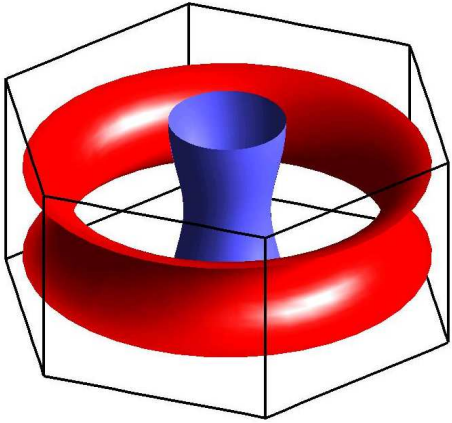}
    \caption{(color). Fermi surface topology used for the calculation of 
     $B_{c2}$ in this work. The $\pi$-band is modeled by a half-torus,
     the $\sigma$-band by a distorted cylinder.
     \label{fig1} }
  \end{center}
\end{figure} 

Usually this eigenvalue problem is
solved by a Landau level expansion of $\Delta_\alpha (\vec{r})$ above
the Abrikosov ground state of the vortex lattice. It has been shown
recently, however, that a variational ansatz for  $\Delta_\alpha (\vec{r})$
corresponding to a {\it distorted} Abrikosov lattice leads to much better
results for strongly anisotropic systems \cite{Posa1} and we will adopt that 
method here. In this method
the ansatz reads $\Delta_\alpha (\vec{r}) = \Delta_\alpha \psi_\Lambda^\tau
(\vec{r})$ with $\psi_\Lambda^\tau (x,y) = \psi_\Lambda (e^{-\tau} x, e^\tau y)$.
Here, $\psi_\Lambda$ is the usual Abrikosov groundstate and $\tau$ is a
variational parameter describing the distortion of the vortex lattice. The
undistorted lattice corresponds to $\tau=0$ and $\tau$ has to be
determined by maximizing $B_{c2}$. Introducing this ansatz into Eqs. (\ref{eq1})
and (\ref{eq2}) and using standard operator techniques (see e.g. Refs. 
\cite{Rieck,Schopohl2,Graser,SunMaki}) we are led to the following
eigenvalue problem for $\Delta_\alpha$ in band space:
\begin{equation}
\Delta_\alpha = 
\sum_{\alpha'} \lambda^{\alpha \alpha'} \left[ \frac{1}{\lambda_+} -
\ln \frac{T}{T_c} - l_{\alpha'} (\tau, \frac{B_{c2}}{T^2} ) \right] \Delta_{\alpha'}
\label{eq3}
\end{equation}
Here, $\lambda_+$ is the highest eigenvalue of $\lambda^{\alpha \alpha'}$,
which determines $T_c$. The function $l_{\alpha}$ is given by the expression
\begin{equation}
l_{\alpha} =
\int_0^\infty \frac{du}{\sinh u} \left\langle 1 -
e^{-u^2 \frac{eB_{c2}}{8 \pi^2 T^2} \left( e^{-2\tau} v_{F 1, \alpha}^2
(\hat{k}) +  e^{2\tau} v_{F 2, \alpha}^2 (\hat{k}) \right) }
\right\rangle_\alpha \label{eq4}
\end{equation}
Here, $v_{F 1, \alpha}$ and $v_{F 2, \alpha}$ are the components of the
Fermi velocity perpendicular to the magnetic field $\vec{B}$ on
Fermi surface $\alpha$. According to Eq. (\ref{eq3}) the criterion for
$B_{c2}$ is that the highest eigenvalue of the matrix 
$\lambda^{\alpha \alpha'} \left[ \frac{1}{\lambda_+} -
\ln \frac{T}{T_c} - l_{\alpha'} (\tau, \frac{B_{c2}}{T^2} ) \right]$
becomes 1. Apparently, at $T=T_c$ this is fulfilled for $B_{c2}=0$,
because $l_{\alpha} \rightarrow 0$.

Eqs. (\ref{eq3}) and (\ref{eq4}) allow to determine the temperature
dependence and angular dependence of $B_{c2}$ from microscopic
grounds. The material parameters we need for the solution are the
Fermi velocities $\vec{v}_{F,\alpha} ( \hat{k} )$ and the
coupling matrix $\lambda^{\alpha \alpha'}$, which can be taken from
bandstructure calculations. In order to simplify the analysis we
restrict ourselves to two relevant bands, because the two $\sigma$-bands 
and the two $\pi$-bands are very similar \cite{Liu}. 
The $\sigma$-band can be described to a good approximation
by a cylindrical Fermi surface with a small $c$-axis hopping parameter.
The $\pi$-band can be modeled by a half-torus as shown in Fig.~\ref{fig1}
(for comparison see the $\Gamma$-point centered Fermi surfaces obtained
from bandstructure calculations in Fig.~1 in Ref.~\cite{Shulga}).
For the Fermi velocities on these two Fermi surfaces we thus write
$\vec{v}_{F,\pi}(\theta,\phi)=
v_{F\pi} ( \cos \theta \cos \phi, \cos \theta \sin \phi, \sin \theta)$
and
$\vec{v}_{F,\sigma}(k_c,\phi)=
v_{F\sigma} (\cos \phi,\sin \phi,\epsilon_c \sin c k_c)$.
Here, $\phi$ is the azimuthal angle within the $ab$-plane, 
$\theta \in [\frac{\pi}{2}, \frac{3\pi}{2}]$ the polar
angle of the torus, $k_c$ the $c$-axis component of the momentum, and $c$ the
lattice constant in $c$-direction. The dimensionless parameter $\epsilon_c$
describes the small $c$-axis dispersion of the cylinder. The Fermi surface
averages over the cylinder and the torus are then given by
$\langle \cdots \rangle_\sigma = \frac{c}{4\pi^2} \int_{-\pi/c}^{\pi/c} dk_c
\int_0^{2\pi} d\phi \cdots$ and
$\langle \cdots \rangle_\pi = \frac{1}{2\pi^2} \int_{\pi/2}^{3\pi/2} d\theta
\int_0^{2\pi} d\phi \frac{1+\kappa \cos \theta}{1-2 \kappa/\pi} \cdots$,
where $\kappa$ is the ratio of the two radii of the torus.
The parameters of the Fermi velocities
can be found from bandstructure calculations. Taking the values given in
Ref.~\cite{Brinkman}, we determine $v_{F\pi}=8.2 \cdot 10^5$m/s, 
$v_{F\sigma}=4.4 \cdot 10^5$m/s, and $\epsilon_c=0.23$. The ratio of the
two radii of the torus can be estimated from the Fermi surface
crossings in Ref.~\cite{Kortus} to be about $\kappa=0.25$. 

For a $2 \times 2$ matrix $\lambda^{\alpha \alpha'}$ the criterion that the
biggest eigenvalue of Eq.~(\ref{eq3}) becomes 1 leads to the
equation
$(1-\eta) l_\sigma + \eta \; l_\pi + \ln t =
- \Lambda_{\pm}
\left( l_\sigma + \ln t \right)
\left( l_\pi + \ln t \right)$.
Here, $t=T/T_c$, $\lambda_-$ is the smaller eigenvalue of $\lambda^{\alpha \alpha'}$
and $\eta=\frac{\lambda^{\pi \pi} - \lambda_-}{\lambda_+ - \lambda_-}$ is
a dimensionless parameter describing the interband coupling strength of the
two bands ($\eta=0$ corresponds to no coupling, $\eta=0.5$ to maximum coupling).
From bandstructure calculations in Ref.~\cite{Liu} the following
effective matrix elements can be obtained: $\lambda^{\sigma \sigma}=0.959$,
$\lambda^{\sigma \pi}=0.222$, $\lambda^{\pi \sigma}=0.163$, and
$\lambda^{\pi \pi}=0.278$. From these we find $\lambda_+=1.008$,
$\lambda_-=0.228$, and $\eta=0.064$.
As it turns out the remaining parameter 
$\Lambda_{\pm}=\frac{\lambda_+ \lambda_-}{\lambda_+ - \lambda_-}$ only weakly affects
the results as soon as $\lambda_-$ is sufficiently smaller than $\lambda_+$,
as is the case here.

\begin{figure}
  \begin{center}
    \includegraphics[width=0.65\columnwidth,angle=270]{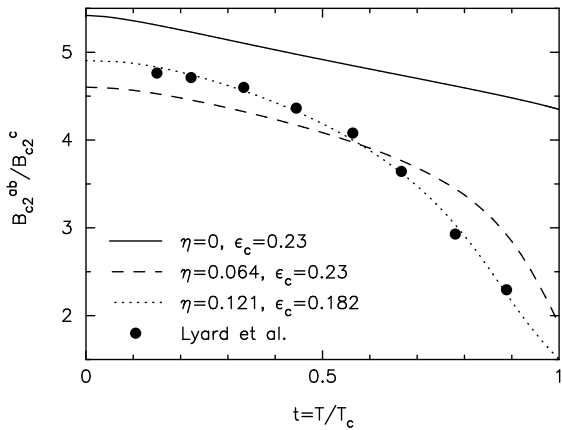}
    \caption{Temperature dependence of the anisotropy ratio
     $\Gamma=B_{c2}^{ab}/B_{c2}^c$ for the two-band model
     described in the text and different interband coupling
     strengths $\eta$. Solid circles are experimental results
     taken from Lyard et al.\cite{Lyard}.
     \label{fig2} }
  \end{center}
\end{figure} 

We have calculated $B_{c2}$ numerically for the material parameters above
optimizing the parameter $\tau$ such that $B_{c2}$ is maximized. 
In Fig.~\ref{fig2}
we show our result for the anisotropy ratio $\Gamma=B_{c2}^{ab}/B_{c2}^c$ as
a function of temperature for different values of the interband coupling strength
$\eta$. For $\eta=0$, when there is no coupling between the two bands, the
temperature dependence of $\Gamma$ is determined by the cylindrical
$\sigma$-band because of its stronger pairing interaction. Here, $\Gamma$ 
changes only by 20 percent, as one expects for an
isotropic single gap superconductor. When $\eta$ is increased, however, the
temperature dependence of $\Gamma$ becomes more pronounced. 
Our result for the parameters given above is shown as the dashed line.
Note, that our calculation is parameter free, relying only on the parameters
given by bandstructure calculations. For comparison, also the experimental
data by Lyard et al. \cite{Lyard} are shown (solid circles). As will become
clear below, the most important parameters determining the temperature dependence
of $\Gamma$ are the interband coupling strength $\eta$ and the $c$-axis
dispersion parameter $\epsilon_c$. If we allow these two parameters to vary
somewhat, we can obtain an excellent fit of the experimental data. The dotted
line shows our result for $\eta=0.121$ and $\epsilon_c=0.182$, where the other
parameters have been kept constant. We mention here that this set of parameters
also gives a correspondingly good fit of the temperature dependences of
$B_{c2}^c$ and $B_{c2}^{ab}$ separately including the upward curvature of 
$B_{c2}^{ab}$ that has been noted in the experiments \cite{Graser2}. 
This is just an immediate consequence of 
the strong temperature dependence of the anisotropy ratio and
$B_{c2}^c$ varying linearly near $T_c$ because of an absence of vortex
lattice distortion in this field direction.

\begin{figure}
  \begin{center}
    \includegraphics[width=0.65\columnwidth,angle=270]{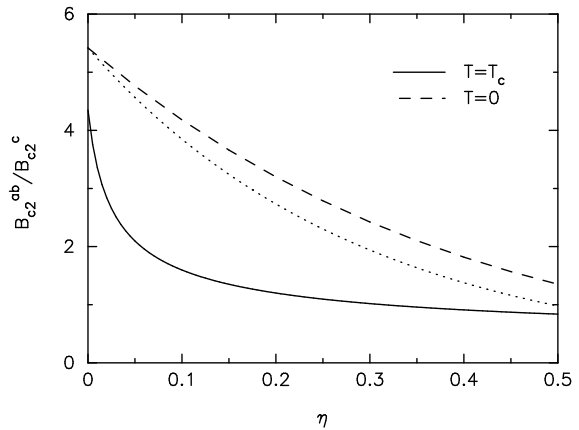}
    \caption{Anisotropy ratio $\Gamma=B_{c2}^{ab}/B_{c2}^c$
     as a function of interband coupling strength $\eta$ for
     $T=0$ (dashed line) and $T=T_c$ (solid line). The dotted line shows
     the approximation for $T=0$ given in Eq.~(\ref{eq10}). At $T_c$ the 
     anisotropy ratio is much
     more sensitive to small interband coupling strengths $\eta$.
     \label{fig3} }
  \end{center}
\end{figure} 

Apparently, the anisotropy ratio $\Gamma$ turns out to be more sensitive to
interband coupling $\eta$ close to $T_c$ than at $T=0$. In order to get a
better physical understanding for this behavior, we want to discuss
some limits, in which analytical solutions for $B_{c2}$ can be obtained.
At first we consider the limit $T \rightarrow T_c$. In this case
Eqs. (\ref{eq3}) and (\ref{eq4}) can be solved exactly and we
obtain for the angular dependence of $B_{c2}$ as a function of the
angle $\beta$ the magnetic field makes with the $c$-axis:
\begin{equation}
\frac{B_{c2}(\beta)}{B_{c2}(\beta=0)} =
(\cos^2 \beta + A  \sin^2 \beta)^{-1/2}
\label{eq6}
\end{equation}
where
$ A = 2
\frac{ \left( 1-\eta \right) 
\langle v_{c,\sigma}^2 \rangle_\sigma +
\eta  \langle v_{c,\pi}^2 \rangle_\pi } 
{\left( 1-\eta \right) \langle v_{ab,\sigma}^2 \rangle_\sigma +
\eta  \langle v_{ab,\pi}^2 \rangle_\pi }
$.
Here, $\langle v_{c,\alpha}^2 \rangle_\alpha$ is the average of the squared
$c$-axis component of the Fermi velocity over Fermi surface $\alpha$, while
$\langle v_{ab,\alpha}^2 \rangle_\alpha$ is the corresponding in-plane
quantity. The distortion parameter $\tau$ is given by $e^{-2\tau}=
\frac{B_{c2}(\beta)}{B_{c2}(\beta=0)}$.
When we introduce the numerical values for $v_{F\pi}$, $v_{F\sigma}$ 
and $\kappa$ given above we find for the anisotropy ratio
\begin{equation}
\frac{B_{c2}^{ab}}{B_{c2}^c} = \frac{1}{\sqrt{A}} =
\frac{1}{\epsilon_c} \sqrt{\frac{1+0.63 \eta}{1+\eta 
\left( \frac{3.69}{\epsilon_c^2}-1 \right)}}
\label{eq8}
\end{equation}
For $\epsilon_c=0.23$ this result is shown in Fig.~\ref{fig3} as
a function of $\eta$ (solid line). From Eq.~(\ref{eq8}) we see that at
$\eta=0$ the anisotropy is given by the $c$-axis dispersion of the
cylindrical Fermi surface and actually diverges for 
$\epsilon_c \rightarrow 0$. (Note, that this divergence could not have been
obtained from a Landau level expansion above the Abrikosov ground state and
our variational ansatz above is crucial for this result).
Once the interband coupling $\eta$ is increased, the
anisotropy quickly reduces. This reduction becomes sizeable already, 
when $\eta \sim \frac{\langle v_{c,\sigma}^2 \rangle_\sigma}
{\langle v_{c,\pi}^2 \rangle_\pi}=\frac{\epsilon_c^2}{3.69}$. Thus,
it is the small $c$-axis Fermi velocity of the cylindrical Fermi surface
as compared to the one of the $\pi$-band which leads to a high sensitivity 
of $\Gamma$ near $T_c$ and a small interband coupling is already sufficient
to make the influence of the $\pi$-band visible.

In the limit $T \rightarrow 0$
the integration in Eq.~(\ref{eq4}) can be performed and we find
$l_{\alpha} (\tau, \frac{B_{c2}}{T^2} ) =
\frac{1}{2} \ln \left( \frac{\gamma e B_{c2}}{2 \pi^2 T^2} \right) 
+ \frac{1}{2} \left\langle \ln  \left( e^{-2\tau} v_{F 1, \alpha}^2
+  e^{2\tau} v_{F 2, \alpha}^2  \right) \right\rangle_\alpha$
where $\ln \gamma= 0.577$ is Euler's constant. Since the pairing
interaction is dominant in the $\sigma$-band, we can employ
two approximations: at first we set $\lambda_-=0$. 
This corresponds to assuming that superconductivity
in the $\pi$-band is completely induced by interband coupling.
As a second approximation we assume that the vortex lattice
distortion $\tau$ is also dominated by the $\sigma$-band. Then
we can find $\tau$ by just optimizing it at $\eta=0$ and then use it
for evaluation at $\eta>0$. With these two approximations the
logarithmic averages in $l_{\alpha}$ can be performed and we
finally get
\begin{equation}
\frac{B_{c2}^{ab} \left( T=0 \right)}{B_{c2}^c \left( T=0 \right)} 
= \frac{1.246}{\epsilon_c}
e^{-\eta \left( 0.482 - 2 \ln \epsilon_c \right)}
\label{eq10} 
\end{equation}
The parameter $\tau$ is found to be $\tau=\frac{1}{2} \ln \epsilon_c$,
which shows that at low temperatures there is no simple relation
anymore between $\tau$ and the anisotropy ratio. Also we mention here
that at low temperatures Eq.~(\ref{eq6}) does not hold anymore.
In Fig.~\ref{fig3} Eq.~(\ref{eq10}) is shown for $\epsilon_c=0.23$ as the
dotted line. In order to demonstrate the quality of these approximations
also the fully numerical result without these approximations is shown
as the dashed line. Apparently, now the anisotropy $\Gamma$ is much
less sensitive to interband coupling and only varies on a scale given
by $\eta \sim 1/\left( 0.482 - 2 \ln \epsilon_c \right)$. The reason for
this is that $\epsilon_c$ now only comes in logarithmically.

\begin{figure}
  \begin{center}
    \includegraphics[width=0.85\columnwidth,angle=0]{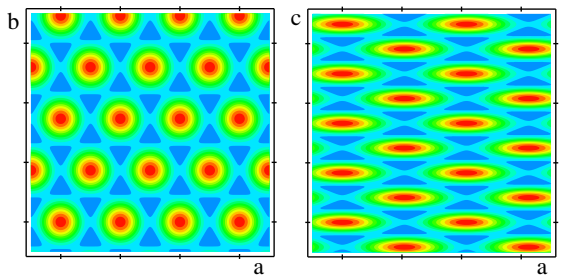}
    \caption{(color). Vortex lattice structure for magnetic field
     in c-axis direction (left panel) and in ab-plane direction 
     (right panel) at zero temperature calculated
     from the two-band model described in the text. For comparison,
     the area of the unit cell has been kept fixed.
     \label{fig4} }
  \end{center}
\end{figure} 

Eq. (\ref{eq10}) shows that at $T=0$ the divergence for 
$\epsilon_c \rightarrow 0$ appears at all $\eta<0.5$, while at
$T=T_c$ this divergence only shows up at $\eta=0$. Numerically
we observe that at finite $\eta$ and $\epsilon_c=0$ there is
a certain temperature at which $B_{c2}^{ab}$ diverges. This
temperature becomes smaller when $\eta$ is increased.
In Fig.~\ref{fig4} we show the distortion of the vortex lattice at
high magnetic field and low temperature that we expect from our calculation. 
When the magnetic field is directed along the $c$-axis of the crystal a
regular Abrikosov lattice is expected as shown in Fig.~\ref{fig4} (left panel).
However, when the field is directed within the $ab$-plane we expect
a distortion of $e^\tau = \sqrt{\epsilon_c} = 0.48$ as shown in 
Fig.~\ref{fig4} (right panel). This prediction can be checked by neutron 
scattering or STM tunneling \cite{Eskildsen}.

To summarize, we have calculated the anisotropy of the upper critical field
for the two band Fermi surface topology shown in Fig.~\ref{fig1}. Using
parameters from bandstructure calculations for MgB$_2$ we find a strong
temperature dependence of the upper critical field in agreement with recent
measurements on MgB$_2$ single crystals. Fine tuning of the parameters can
yield a very good fit of the experimental data. We observe that 
the small $c$-axis dispersion of the $\sigma$-band leads
to a high sensitivity of the anisotropy ratio on the interband coupling 
strength near $T_c$, but not at $T=0$. This suggests that the interplay
of these two quantities leads to the strong temperature dependence 
of the upper critical field in MgB$_2$.

We would like to thank F. Bouquet, O.V. Dolgov, M.R. Eskildsen, 
K. Maki, A.I. Posazhennikova and L. Tewordt for valuable discussions.
Thanks are also due to S. Graser for his help.

\end{document}